\documentclass[sigconf, nonacm, authorsversion]{acmart}

\usepackage{booktabs} 
\usepackage{makecell}

\usepackage{todonotes}

\setcopyright{none}

\acmConference[]{}{}{}

\acmYear{2019}
\copyrightyear{2019}
\begin{document}
\title{Enhanced Performance for the encrypted Web\\through TLS Resumption across Hostnames}

\author{Erik Sy}
\affiliation{%
  \institution{University of Hamburg}
}

\author{Moritz Moennich}
\affiliation{%
  \institution{University of Hamburg}
}

\author{Tobias Mueller}
\affiliation{%
  \institution{University of Hamburg}
}

\author{Hannes Federrath}
\affiliation{%
  \institution{University of Hamburg}}

\author{Mathias Fischer}
\affiliation{%
  \institution{University of Hamburg}}
\renewcommand{\shortauthors}{Sy et al.}

\begin{abstract}

TLS can resume previous connections via abbreviated resumption handshakes that significantly decrease the delay and save expensive cryptographic operations.
For that, cryptographic TLS state from previous connections is reused.
TLS version~1.3 recommends to avoid resumption handshakes, and thus the reuse of cryptographic state, when connecting to a different hostname.
In this work, we reassess this recommendation, as we find that sharing cryptographic TLS state across hostnames is a common practice on the web.
We propose a TLS extension that allows the server to inform the client about TLS state sharing with other hostnames.
This information enables the client to efficiently resume TLS sessions across hostnames.
Our evaluation indicates that our TLS extension provides huge performance gains for the web. For example, about 58.7\% of the 20.24 full TLS handshakes that are required to retrieve an average website on the web can be converted to resumed connection establishments.
This yields to a reduction of 44\% of the CPU time consumed for TLS connection establishments.
Furthermore, our TLS extension accelerates the connection establishment with an average website by up to 30.6\% for TLS 1.3. Thus, our proposal significantly reduces the (energy) costs and the delay overhead in the encrypted web.
\end{abstract}

%
%



\maketitle

\section{Introduction}

To communicate securely, the web increasingly adopts transport encryption via TLS~\cite{felt2017measuring}.
Nowadays, more than 75\% of all web requests are protected using TLS encryption~\cite{HTTP_Archive}.
This practice benefits communication security, but leads to additional performance penalties.
Most web transactions are short transfers that are significantly delayed by the TLS connection establishment.

To accelerate the connection establishment, TLS~1.3~\cite{rfc8446} and its predecessors provide session resumption mechanisms.
They abbreviate the TLS handshake based on a shared secret exchanged during a prior TLS session between the client and server.
In total, these resumption handshakes significantly reduce computational overhead for cryptographic operations and save up to one round-trip between client and server.

Load balancing can make it necessary to provide the content of a hostname via different servers.
To sustain the benefits of TLS resumptions in such a setup, the respective servers can share their secret TLS state related to the client's connection.
Furthermore, TLS~1.3~\cite{rfc8446} allows resumption handshakes across hostnames when they share the same TLS certificate.
Thus, the client can resume a previous connection to hostname $\mathcal{A}$ with hostname $\mathcal{B}$.
However, TLS~1.3~\cite{rfc8446} recommends not to use TLS resumption across hostnames.
We find that this recommendation of TLS~1.3 leads to a significant performance limitation.

As an illustration for such a performance limitation, we assume a website served from the hostname \textit{www.example.com}.
Moreover, we assume that the hostname \textit{example.com} is operated by the same entity and provides a redirect to \textit{www.example.com}.
A client that retrieves the website via the hostname \textit{example.com} thus needs to establish two TLS connections.
Following the specification of TLS~1.3, the client is required to establish both connections with two full handshakes.
However, as both hostnames are operated by the same entity we assume that they can share their secret TLS state with each other.
Thus, a performance optimized website retrieval requires only a full handshake to \textit{example.com} and subsequently conducts a resumption handshake to connect to \textit{www.example.com}.

In this work, we propose and evaluate a TLS extension that aims to enable the client to efficiently resume TLS connections across hostnames.
In summary, this paper makes the following contributions:
\begin{itemize}
\item we introduce a TLS extension that enables the efficient use of resumption handshakes across different hostnames.
\item we simulate the loading behavior of popular websites to assess the performance impact of our proposal.
Our results indicate, that our approach can convert about 58.7\% of the required full TLS handshakes to performance optimized resumption handshakes upon the first visit of an average website.
This reduces the computational costs for the respective TLS handshakes by about 44\% and accelerates the establishment of TLS~1.3 connections by up to 30.6\%.
\end{itemize}

The remainder of this paper is structured as follows: Section~\ref{sec:Problem} briefly introduces TLS resumption and describes the performance problems of TLS resumption per hostname that we aim to solve. 
Section~\ref{sec:Delegation} summarizes the proposed TLS extension.
The evaluation results are presented in Section~\ref{sec:Evaluation}. 
Security and privacy considerations of the proposed TLS extension are discussed in Section~\ref{sec:Discussion}, and related work is reviewed in Section~\ref{sec:Related}.
Section~\ref{sec:Conclusion} concludes the paper.

\section{Problem Statement} \label{sec:Problem}
In this section, we review the session resumption mechanism as known from TLS 1.3~\cite{rfc8446}.
Then, we describe the performance problem of TLS resumption per hostname that we aim to solve.

\subsection{TLS resumption}

Transport Layer Security (TLS) is a cryptographic protocol that provides authentication, confidentiality, and data integrity for end-to-end communication.
It finds widespread use in applications such as web browsing, email, and voice over IP (VoIP).
The latest version of it is TLS 1.3~\cite{rfc8446}.
It provides the option to establish an encrypted channel with either a full or a resumed handshake.
Compared to the full handshake, the resumed handshake of TLS~1.3 provides significant performance gains.
The most time-consuming operations in a TLS full handshake are related to the authentication of the server's identity~\cite{bench_wolfssl}.
These operations involve the computation of a signature with the server's private key.
Additionally, they require the client to verify the server's certificate, which includes to check the full certificate chain and to verify the server's signature with the public key contained in the presented certificate.

Resumed handshakes reuse the cryptographic state of a previous connection between two communication partners.
In the resumption handshake, the communication partners authenticate each other by their ability to reuse the cryptographic state of a previous connection. 
As a result, resumption handshakes skip the computationally expensive public-key cryptography to authenticate the server's identity.
Besides a lower computational overhead for resumed connections, TLS~1.3 provides also a resumed handshake that requires one round-trip time (RTT) less than the full handshake to establish the encrypted channel.
As a drawback, these resumed TLS~1.3 handshakes impact the client's privacy as the shared secret can be used as a tracking mechanism~\cite{sy2018tracking}.
Furthermore, the resumed 0-RTT handshake does not protect the server's applications against replay attacks, which TLS itself guards against for other handshake modes.
For a comprehensive description of this reduced security guarantees of the 0-RTT resumed handshake and more details on TLS~1.3, we refer readers to RFC~8446~\cite{rfc8446}.

\subsection{The limitation of TLS resumption per SNI}

According to the specification of TLS 1.3~\cite{rfc8446}, a resumption handshake \textit{should} only be used when the hostname of the server, also known as server name indication (SNI), matches the server's hostname of the original TLS session.
Because of this recommendation of TLS~1.3, the use of resumption handshakes is practically restricted to revisits to servers with the same SNI.
This approach provides only performance benefits during the first visit of a website if it is required to establish multiple TLS connections to the same SNI.

However, we find that the web has a complex structure where accessing a web page requires an average of 19 TCP connections to several hosts~\cite{HTTP_Archive}.
Assuming that all of these connections are TLS encrypted, this results in 19 TLS full handshakes for the first visit of an average website.
TLS resumption across hostnames can potentially further reduce this large number of TLS full handshakes to realize performance improvements.

Only with respect to network latency, these TLS full handshakes can cause a significant latency overhead \textit{x} during web page loading compared to the use of resumed 0-RTT handshakes.
Note, that each resumed 0-RTT handshake saves one round-trip time compared to the full handshake.
For the 19 connections of the average website, this reduced latency overhead can sum up to 19 RTTs when all connections are established sequentially.
The lower boundary of the reduced latency overhead is one RTT if all 19 connections are established in parallel.
Equation~\ref{eq:latency_overhead} shows these boundaries of the latency overhead \textit{x}.
\begin{equation}\label{eq:latency_overhead}
RTT \leq x \leq \sum_{i=0}^{19} RTT_{i}
\end{equation}
The average round-trip time for mobile LTE connection in the U.S. is approx. 60\,ms to reach popular online services~\cite{opensignal}.
Thus, for the considered average web page the induced latency overhead by these initial handshakes can reach up to 1.14\,s for LTE connections.
Note, that the round-trip time for 3G and WiFi connections are on average longer than for LTE connections, with a latency of 212~ms and 151~ms respectively~\cite{opensignal2}.
Therefore, these connections types experience an even longer latency overhead for the same web page.  

However, the use of resumed handshakes also significantly reduces the latency overhead with respect to the CPU time.
Measurements of the CPU time for TLS 1.2~\cite{cloudflare_CPU} indicate that a resumed handshake requires only 0.3~ms compared to a full handshake with 6.9~ms.
Thus, based on this measurement the computational effort of a full handshake seem to be 23 times higher than for a resumed handshake.
With respect to the latency overhead, each resumed connection saves about 6.6~ms in CPU time on the sampled test setup.
For the average website, this leads to savings in the page loading time between 6.6~ms for parallel connections and 125.4~ms for 19 sequential connections.


In total, the above paragraphs highlight that the latency overhead of loading web pages can be significantly reduced if TLS full handshakes are replaced with resumed handshakes.
However, the restriction of TLS~1.3 to use resumption handshakes only for revisits to the same SNI provides a practical barrier to increase the number of resumed handshakes during the first visit of a website.
For example, a redirect from \textit{https://example.com} to \textit{https://www.example.com} requires two TLS full handshakes because these are two different SNIs.
Similar, if \textit{https://www.example.com} includes resources from \textit{https://static.example.com} the loading of the website requires two independent TLS full handshakes. 
This policy of TLS~1.3 is appropriate if it is intended that a TLS resumption handshake between the different SNIs \textit{example.com}, \textit{www.example.com}, and \textit{static.example.com} should not be possible.
For example, if \textit{www.example.com}, and \textit{static.example.com} are operated by different entities that do not trust each other.
However, if  \textit{www.example.com}, and \textit{static.\-example.com} are operated by the same entity or by entities that trust each other, then resumption handshakes between these SNIs should be possible.
Hence, this approach would use resumption handshakes with SNIs, that have not been visited previously.
As a result, the latency overhead from loading a website that includes resources from many SNIs for the first time can be significantly decreased by this approach, when some of these SNIs have a trust relationship between each other.  

Note, that the specification of TLS~1.3~\cite{rfc8446} appreciates that performance optimizations are feasible if clients resume to servers with different hostnames.
However, it does not provide an approximation of these performance gains nor does it describe means to implement the exploitation of these performance optimizations.

\section{TLS resumption across hostnames} \label{sec:Delegation}

In this section, we propose a TLS extension that guides the use of TLS resumption across different hostnames, i.e., it allows to inform clients that a TLS connection with SNI $\mathcal{A}$ can also be resumed with SNI $\mathcal{B}$.
This information can be forwarded during the TLS handshake via our proposed TLS extension with the type \textit{resumption\_across\_sni}.

Clients that support the TLS extension can indicate this within their \textit{ClientHello} message.
For this purpose, they include the appropriate extension type with an empty data field to their message.
Servers must only send the \textit{resumption\_across\_sni} extension to clients who signaled their support for it within their \textit{ClientHello}.
In this case, the server sends a list of SNIs that support the resumption based on the TLS state of the original session.
This list of SNIs is sent within the data field of the \textit{resumption\_across\_sni} extension, which in turn is sent as extension data for the servers certificate but is not part of the certificate itself. 
Note, that this part of the server's response uses transport encryption to protect against network-based attackers.  

Upon receiving the server's response, the client must verify that each SNI provided within the \textit{resumption\_across\_sni} extension is also contained in the server certificate. 
The specification of TLS~1.3~\cite{rfc8446} instructs that a client can only resume a session with a new SNI $\mathcal{B}$, if the server certificate presented in the original session with SNI $\mathcal{A}$ is valid for SNI $\mathcal{B}$. 
Thus, the client should only use the session resumption mechanism with SNIs that can be authenticated by the private key of the server's TLS certificate.

To establish a connection to one of these SNIs, the client can directly use a resumption handshake for which it utilizes the state of the original session.
To support the proposed TLS extension on the server-side, the involved SNIs are required to share cryptographic state among each other. For TLS~1.3 this depends on the construction approach of the \textit{pre-shared keys} that are used to conduct the resumption of a prior session.
The encrypted TLS connection state either directly contains a \textit{pre-shared key} or the connection state contains a database lookup that refers to the \textit{pre-shared key} in a database instead.
In the direct case, the involved SNIs need to share a secret key that enables them to retrieve the secret server-side connection state from the \textit{pre-shared key} provided by the client.
In the database case, the different SNIs need to share their access to a common database containing their secret server-side connection states. 

Note, that the proposed TLS extension can be used in combination with TLS version~1.2 and lower.
Compared to TLS~1.3, these older TLS versions do not apply the concept of encrypted extensions.
Hence, the server's response has to be transmitted in the unencrypted list of extensions within the respective \textit{ServerHello} message.

\section{Evaluation} \label{sec:Evaluation}

In this section, we investigate the real-world impact of our proposed TLS extension.
For that, we first study the delay overhead of different TLS handshakes to substantiate the performance benefits of TLS resumption. Subsequently, we investigate real-world websites and the number of sequential TLS connections required to load them as well as existing trust relationships in between hostnames within the respective domain trees.
Finally, we simulate the performance impact of our proposed TLS extension on the page loading behavior of the investigated websites.

\subsection{Delay overhead of TLS connection establishment}\label{sec:delay_overhead}

 \begin{table*}[tb]
 \caption{Mean duration to establish a TLS connection via different handshake modes between the client-server pair and to conduct a short data transfer.}
  \label{tab:Resumption_comparison}
 \centering
 \begin{tabular}{lrrrrr}
 \toprule
\multicolumn{1}{c}{Network} & \multicolumn{2}{c}{TLS 1.2} & \multicolumn{3}{c}{TLS 1.3} \\
 \cmidrule(lr){2-3}
 \cmidrule(lr){4-6}
\multicolumn{1}{c}{latency} & \multicolumn{1}{c}{Full} & \multicolumn{1}{c}{Resumed} & \multicolumn{1}{c}{Full} & \multicolumn{1}{c}{1-RTT resumed}& \multicolumn{1}{c}{0-RTT resumed} \\
 \midrule
$\approx$0.3 ms& 26.66 ms & 2.69 ms & 29.17 ms  & 6.34 ms & 6.57 ms \\
50 ms& 237.86 ms & 154.20 ms & 190.06 ms  & 160.12 ms& 109.61 ms  \\
100 ms & 439.08 ms & 304.50 ms & 340.81 ms & 310.27 ms& 209.72 ms \\
150 ms& 639.15 ms & 454.621 ms  & 490.87 ms & 460.26 ms & 309.44 ms \\
 \bottomrule
 \end{tabular}
 \end{table*}

In this section, we study the delay overhead and the CPU time of different TLS handshake modes. The delay overhead is the additional time experienced by a user if a full TLS handshake instead of a resumed handshake is used.
The CPU time reflects more an economic and environmental perspective, as a CPU is capable to resume more connections than conducting full TLS handshakes within the same time frame. 

\subsubsection{Evaluation setup}\label{sec:eval_setup} 

For the experiment, we compare the required time to download a small web page from a single host using standard TCP together with different TLS handshake versions to establish the encrypted channel. The tested TLS handshakes are full TLS~1.2, TLS~1.2 resumption via session identifier, full TLS~1.3, 1-RTT TLS~1.3 resumption, and 0-RTT TLS~1.3 resumption.

We use two virtual machines, one acting as web server and the other one as client.
The virtualization is realized on the same host using  qemu 2.8 and libvirt 3.0.0.
This test setup leads to short network latencies with an average ping of 0.3 milliseconds (ms) between the virtual machines. The host machine was equipped with an Intel Xeon E5-1660 v4 CPU with 32GB of RAM and runs Debian stretch. The client and server virtual machines were both set up with 4~GB of RAM and were running an Ubuntu 18.10, respectively. The server ran the example server program shipped with the wolfSSL library~\cite{wolfssl}.
After successfully establishing a TLS connection, this program responds with a short string to the client's request.
The client ran the corresponding example client program of wolfSSL that establishes a TLS session to the server, issues a request and waits for the server's response before it terminates the TLS session.
 
For TLS~1.3 we used the forward-secure cipher suite \textit{TLS\_AES\_128\-\_GCM\_SHA256}.
As this particular cipher suite is not available in TLS~1.2, we used the most similar forward-secure cipher suite \textit{ECDHE\_RSA\_AES128\-\_GCM\_SHA256}.
To account for skews in the measurements, we repeated the experiment 1000 times and measured the elapsed wall-clock time.
We conducted our measurements with the client's network interface configured to simulate network latencies of 0.3~ms, 50~ms, 100~ms, and 150~ms with iproute2's \texttt{tc} program.
We recorded and inspected the network traffic of the virtual network interface to validate a correct behavior of our evaluation setup.

\subsubsection{Measuring the elapsed time}\label{sec:elapsed_time}

The delay overhead measurements are summarized in Table~\ref{tab:Resumption_comparison}.
We find that independent of the tested TLS version and network latency the resumed connection establishment saves at least 22.6~ms compared to the full handshakes.
For TLS~1.2 and a latency of approximately 0.3~ms this reduces the overhead of the connection establishment by a factor of almost ten.
The absolute reduction of the delay overhead between the full handshake and resumed connection establishment increases with an increasing network latency, except for the TLS~1.3 1-RTT resumed handshake.
This behavior can be attributed to the saved round-trip times from resumed TLS~1.2 and resumed TLS~1.3 0-RTT handshakes compared to the corresponding full handshake.
We observe, that a TLS~1.2 full handshake requires one RTT delay to establish the transport connection, another two RTTs to establish the encrypted channel, before the last RTT requests and retrieves the small web page.
As expected, the TLS~1.3 full handshake abbreviates this TLS connection establishment by a single RTT.

Considering the number of required round-trips times for each measurement, we find that the duration of the connection establishment is shorter for a network latency of approximately 0.3~ms than for higher network latencies.
For example, the full TLS~1.3 handshake has a delay overhead of one RTT for the TCP connection, one RTT to establish the encrypted channel,  and another RTT to request and retrieve the small web page.
Thus, the delay overhead of the data transfer is about 0.9~ms for a network latency of approximately 0.3~ms and 150~ms for a latency of 50~ms.
Subtracting this delay overhead from the respective results of Table~\ref{tab:Resumption_comparison}, leads to a gap of about 10~ms as shown in Equation~\ref{eq:gap}.
\begin{equation}\label{eq:gap}
190.06~ms-150~ms-29.17~ms-0.9~ms = 9.99~ms
\end{equation}
We assume that this time gap is caused by parallel operations of both peers that occur when the latency in between them is smaller than the time required to compute the necessary cryptographic operations. 
Thus, with higher latencies peers will sequentially compute operations during connection establishment, while shorter network latencies induce overlapping times for cryptographic computations at client and server.

We define the reduced delay by using a 1-RTT resumed handshake or a 0-RTT resumed handshake instead of a full TLS~1.3 connection establishment as $\Delta_{1RTT}$ and $\Delta_{0RTT}$, respectively.
Based on our measurements, we find that $\Delta_{1RTT}$ is in an interval between 22.83~ms and 30.61~ms (see Equation~\ref{eq:delta_1rtt}).
\begin{equation}\label{eq:delta_1rtt}
22.83~ms \leq \Delta_{1RTT}  \leq 30.61~ms
\end{equation}
However, for $\Delta_{0RTT}$ such an interval depends on the round-trip time between the client and server, as shown in Equation~\ref{eq:delta_0rtt}.
Thus, for larger network latencies the performance benefit of using a 0-RTT resumed handshake instead of a 1-RTT resumed connection establishment increases.
\begin{equation}\label{eq:delta_0rtt}
22.3~ms + RTT \leq \Delta_{0RTT}  \leq 31.43~ms + RTT
\end{equation}
For an average LTE connection in the U.S. with a network latency of 60~ms~\cite{opensignal}, we thus expect that a TLS~1.3 connection using a 0-RTT handshake instead of a full handshake reduces the delay overhead at least by 82.3~ms. 

\subsubsection{Measuring the CPU time}

For this measurement, we deployed the same test setup as described in Section~\ref{sec:eval_setup}. Furthermore, we utilized the same example programs of the wolfSSL library~\cite{wolfssl} to run the server and client. However, we started these programs with the \textit{time} program, which measures the CPU time that was consumed during the execution of another program. Thus, it provides us a metric to assess the conducted computations on the client- and server-side during the corresponding TLS connection establishment.

Table~\ref{tab:CPU_time_comparison} provides the mean values of the CPU time for the different TLS versions based on 10\,000 TLS connection establishments each.
A full TLS handshake consumes between 7.84~ms and 9.22~ms of CPU time per peer.
A resumed TLS~1.2 handshakes requires a CPU time between 0.76~ms and 1.33~ms.
Resumed connection establishments with TLS~1.3 require between 2.33~ms and 2.62~ms of CPU time per peer.
Note, that the measured resumed TLS~1.3 handshakes conduct a Diffie-Hellman key exchange, which establishes a forward-secure communication channel.
This forward-secrecy is not provided by resumed TLS~1.2 handshakes, which leads to a smaller CPU time compared to resumed TLS~1.3 connections.    

In total, our results indicate that with respect to the CPU load at least six resumed TLS~1.2 connections can be established instead of a full TLS~1.2 handshake.
For TLS~1.3, around three resumed handshakes can be performed with the same CPU time as a full handshake.
Thus, an increased usage of resumed handshakes on the web allows to significantly reduce the required CPU time of TLS connection establishments. Note, that savings with respect to the CPU time provide various benefits, like a reduced energy consumption of the peers and reduced hardware requirements.   
 
  \begin{table*}[tb]
 \caption{Mean CPU time to establish a TLS connection via different handshake modes between the client-server pair and to conduct a short data transfer.}
  \label{tab:CPU_time_comparison}
 \centering
 \begin{tabular}{lrrrrr}
 \toprule
\multicolumn{1}{c}{Peer} & \multicolumn{2}{c}{TLS 1.2} & \multicolumn{3}{c}{TLS 1.3} \\
 \cmidrule(lr){2-3}
 \cmidrule(lr){4-6}
\multicolumn{1}{c}{} & \multicolumn{1}{c}{Full} & \multicolumn{1}{c}{Resumed} & \multicolumn{1}{c}{Full} & \multicolumn{1}{c}{1-RTT resumed}& \multicolumn{1}{c}{0-RTT resumed} \\
 \midrule
Server& 8.02 ms & 1.33 ms & 7.84 ms  & 2.33 ms & 2.62 ms \\
Client& 8.26 ms & 0.76 ms & 9.22 ms  & 2.38 ms & 2.46 ms  \\
 \bottomrule
 \end{tabular}
 \end{table*}
 
%
%
%
%

\subsection{Investigating the loading behavior of the Alexa Top~1K Sites}\label{sec:loading_behavior}

\begin{figure}[tbp]
\centering
\includegraphics[width=0.47 \textwidth]{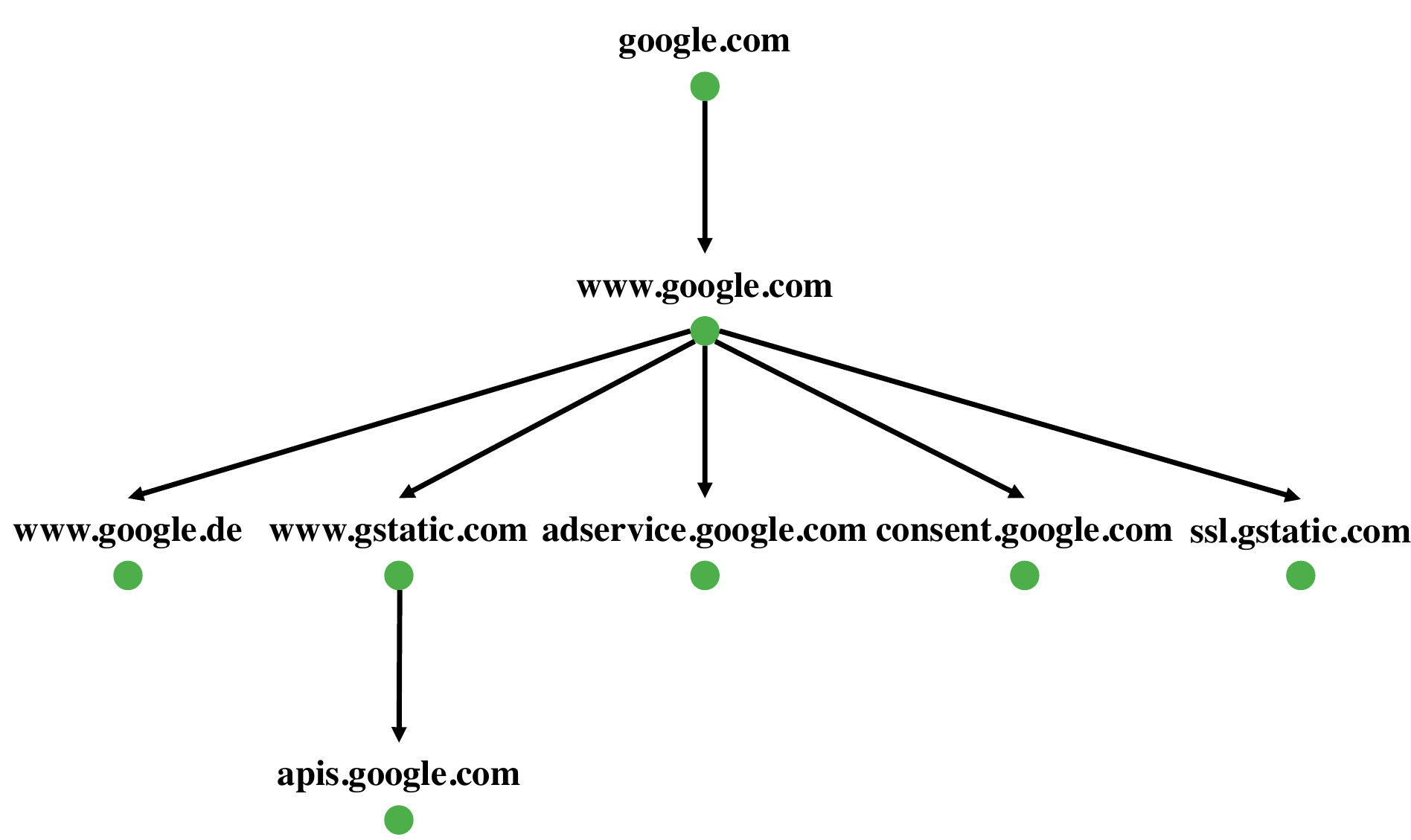}
\caption{Domain tree of root-domain google.com, which holds the first rank within the Alexa Top1K Sites.}
\label{fig:Domain_tree}
\end{figure}

Performance improvements of the proposed TLS extension over standard TLS resumption are possible if a visited hostname shares its TLS state with another hostname that the client will connect to later on. 
Hence, TLS resumption across hostnames can provide performance improvements within the visit of a single website and in between visits of different websites.
In the following, we will investigate the benefits of our proposal with respect to a single visit of a website. 
For this purpose, we derived domain trees for the Alexa Top~1K Sites~\cite{Alexa}.
These domain trees indicate the causal relations between the hostnames requested during loading the respective page.
Figure~\ref{fig:Domain_tree} shows such a domain tree for the root-domain \textit{google.com}.

To generate domain trees, we analyzed results of the online service \textit{urlscan.io}. 
It scans websites with the Google Chrome browser in headless mode and provides an overview of the browser's requests during the page loading.
Furthermore, this service also reports the origin of a new request that allows us to derive the causal sequence of those requests.
We conducted our scans of the Alexa Top~1K Sites on the 8th of November 2018.
We successfully retrieved the domain trees for 839 websites and experienced errors for the remaining 161 domain trees, which manifests an error rate of 16.1\%.
Errors occurred for various reasons on protocols like IP, DNS, SPDY, and TLS.
For example, due to timeouts, unreachable IP addresses, a failed domain name resolution, and protocol errors.
In the following, we assume that the successfully collected website data is qualified to represent the average loading behavior of the Alexa Top~1K Sites.

\begin{figure}[tbp]
\centering
\includegraphics[width=0.47 \textwidth]{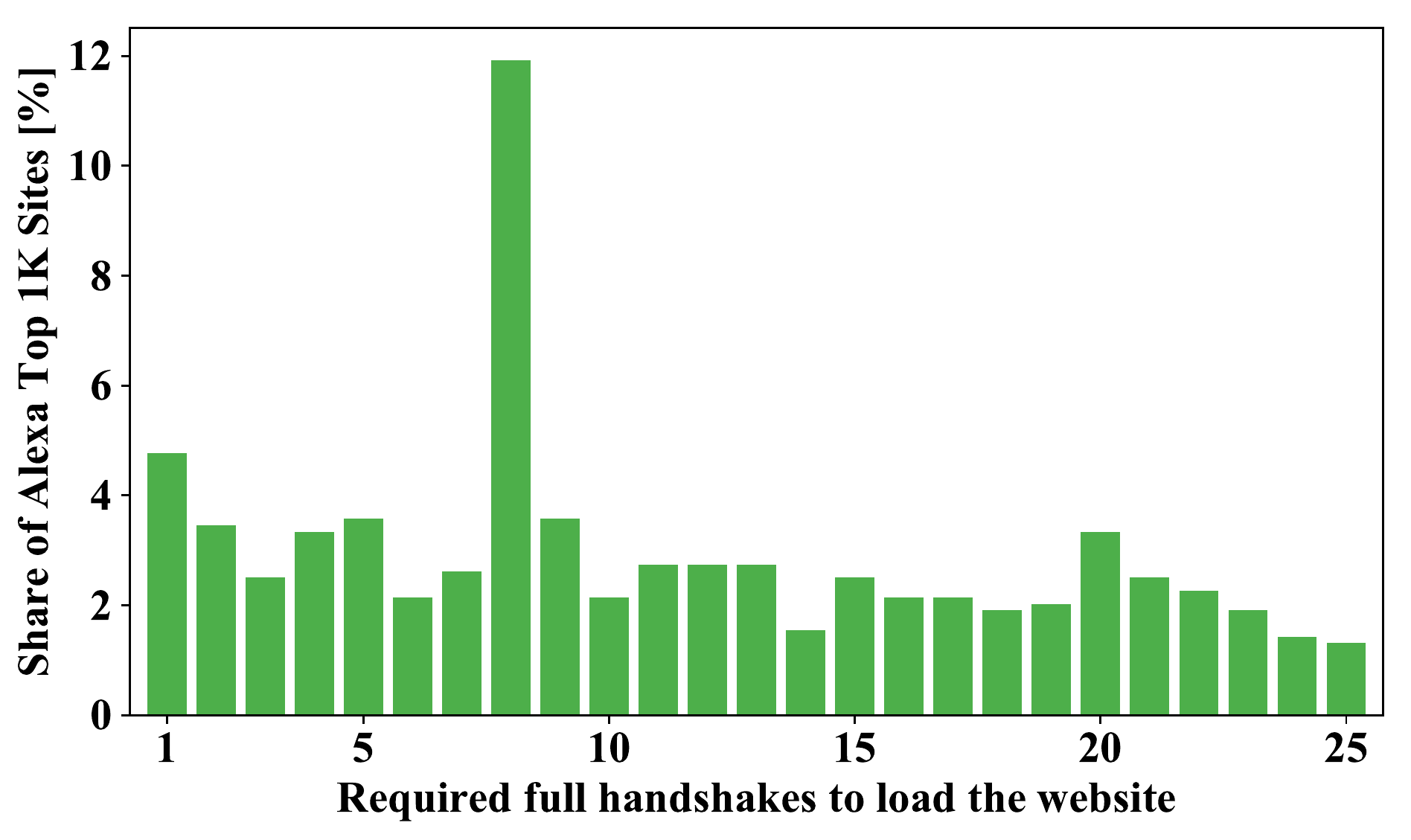}
\caption{Share of Alexa Top~1K Sites over the number of required full TLS handshakes to retrieve the website. Note, that this plot is cut off at 25 full handshakes.}
\label{fig:required_handshakes}
\end{figure}

Based on the collected data, we find the 839 websites required connections to 17\,525 different hostnames to be loaded.
About 97\% of these different hostnames supported TLS encryption.
As our evaluation focuses on the performance improvements for TLS connections, we excluded the remaining 541 hostnames without support for TLS from our evaluation.
On average, we observe that each website within the Alexa Top~1K Sites requires TLS connections to 20.24 different hostnames for its retrieval.
Figure~\ref{fig:required_handshakes} shows the distribution of the required number of TLS connections to different hostnames for the websites in our data set.
We find, that 95.2\% of the investigated websites require more than a single TLS connection for its retrieval. Furthermore, we note that 73.3\% of the analyzed websites require less than 26 full TLS handshake to load the site.
We find that 11.9\% of all observed websites require eight full TLS handshakes, which is the most common configuration. 
Note that also \textit{google.com} utilizes this configuration as shown in Figure~\ref{fig:Domain_tree}.

Using the generated domain trees, we determine the shortest path between the root-domain and all other hostnames within the domain tree.
The longest determined path indicates the maximum number of sequentially established TLS connections required to load the page.
For example, the domain tree of \textit{google.com} requires four sequential TLS connection to load the website (see Figure~\ref{fig:Domain_tree}).
This longest path traverses the hostnames \textit{google.com},  \textit{www.google.com},  \textit{www.gstatic.com}, and  \textit{apis.google.com}.
We use the longest path of sequential TLS connections as metric to describe the impact of TLS connection establishment on the website loading performance.
We assume that a website requires all its TLS connections to be established before the loading of the site can be completed.
For the loading of \textit{google.com}, this leads to four times the delay overhead of a TLS full handshakes until all required TLS connection can be established.

\begin{figure}[tbp]
\centering
\includegraphics[width=0.47 \textwidth]{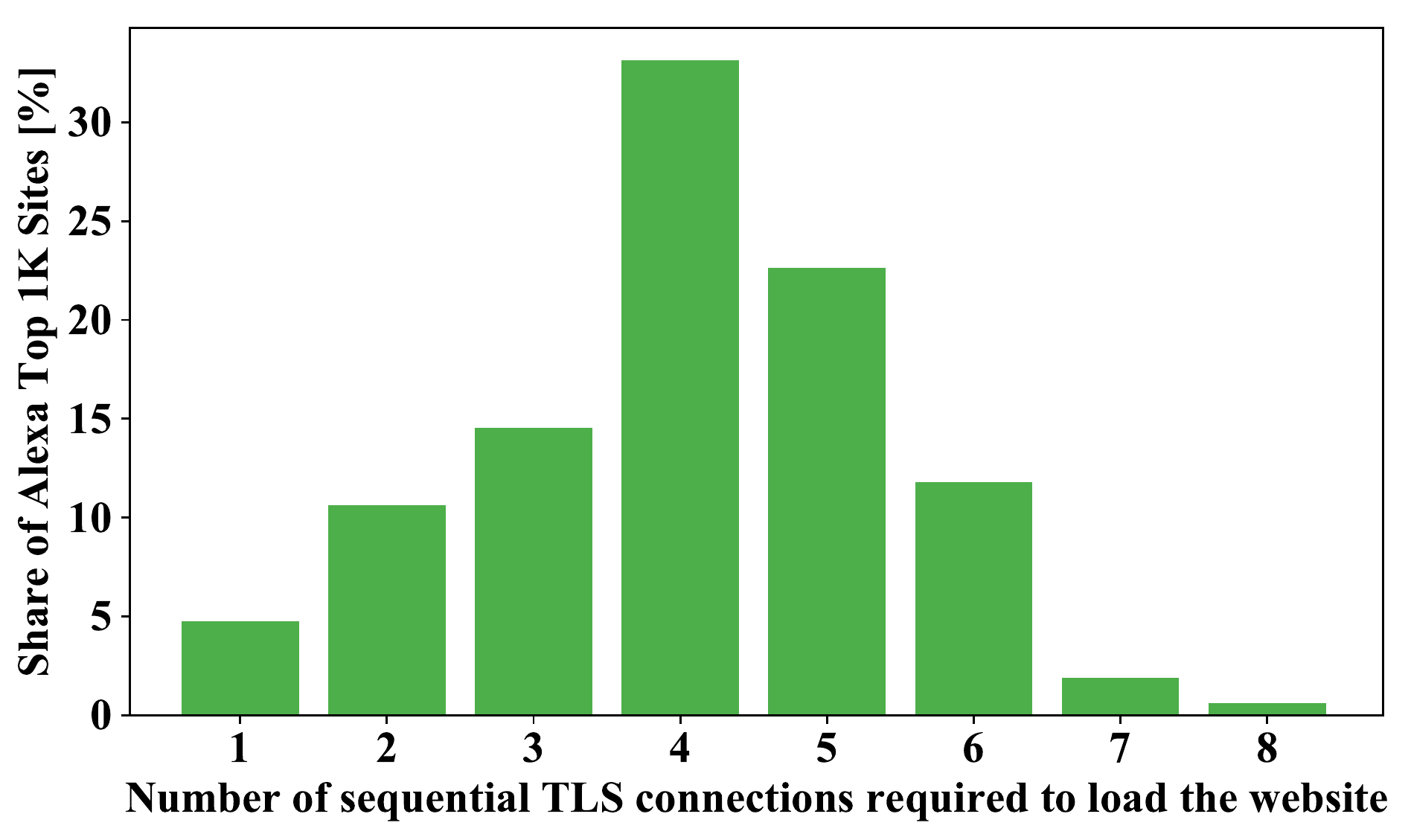}
\caption{Share of Alexa Top~1K Sites is plotted over the number of required sequential TLS connections to load the respective website.}
\label{fig:longest_fullhandshake_path}
\end{figure}  

Figure~\ref{fig:longest_fullhandshake_path} presents a distribution of the investigated websites over the number of sequential TLS connections required loading the site.
As expected from Figure~\ref{fig:required_handshakes}, we find that 4.8\% of the investigated websites require only one sequential TLS connection to retrieve the site.
63.0\% of the websites within the Alexa Top~1K Sites require less than five sequential TLS connections to load them, while no website requires more than eight sequential TLS connections.

\subsection{Measuring trust-relations within the domain trees of the Alexa Top~1K Sites} \label{sec:trust_relations}

To successfully conduct TLS resumption across hostnames, we require the involved hosts to share their secret TLS state. Within this evaluation, we define the sharing of secret TLS state between different hostnames as a trust-relation. We use two approaches to assess trust relations between different hostnames.
First, we assume a trust-relation exists, if a private key of a TLS certificate is valid to authenticate both hostnames.
Second, if a TLS session can be successfully resumed across two hostnames, then we also assume a trust-relation between two hostnames.

To identify trust relations between hostnames based on TLS certificates, we analyze our data collection described in Section~\ref{sec:loading_behavior}.
The results of the Alexa Top~1K Sites include the \textit{Subject Alternative Names (SAN)} stated in the respective TLS certificate.
This SAN-list indicates the hostnames that can be authenticated via the presented certificate and thus can be used to infer trust relations.

We analyzed the SAN-lists of the 16\,984 hostnames supporting TLS that we identified during the scan of the Alexa Top~1K Sites.
For each domain tree, we grouped all hostnames that have a trust-relation with each other.
In total, we find that on average an Alexa Top1K Site connects to 20.24 different hostnames, which form 9.49 trust groups on average.
The mean size of a trust groups is 2.13 as shown in Table~\ref{tab:trust-groups}.
This result indicates that trust relations between nodes of domain trees are a common phenomenon on the web.
However, these determined trust relations do not assure that TLS resumption across different hostnames are feasible.

 \begin{table}[htbp]
   \caption{Mean size of TLS trust groups within the domain trees of each Alexa Top~1K Site}\label{tab:trust-groups}
 \centering
 \begin{tabular}{ccc}
 \toprule
 \makecell{Results based upon\\TLS certificates} & \makecell{Results based upon\\TLS resumption}& \makecell{Union of both\\evaluations}\\
 \midrule
 2.13 & 1.50 & 2.42 \\
 \bottomrule
 \end{tabular}
 \end{table}

To further substantiate the feasibility of our proposal, we attempted to resume TLS sessions between all nodes of each domain tree.
We conducted this measurement on the 25th of January 2019 using the OpenSSL library version~1.1.0f~\cite{openssl} to establish a TLS connection to each node of a respective domain tree.
For this measurement, we used the resumption mechanisms session ID's and session tickets to resume a connection via TLS versions 1.0, 1.1, and 1.2.
We observed that with an exception of 54 hostnames all other hostnames preferred connections via TLS~1.2. Furthermore, our results indicate session tickets are the preferred resumption mechanism for 81.2\% of the hostnames.

For this experiment, we assume that websites supporting the latest TLS version 1.3 will also support one of the earlier TLS versions.
Subsequently, we attempted to resume each initial connection to the domain at each of the other hostnames within the domain trees.
This approach provided us with a list of trust relations for each node of a domain tree.
We evaluated these lists similar to the SAN-lists to determine trust relations within each domain tree of the Alexa Top~1K Sites.

Based upon this measurement, we find that each Alexa Top~1K Site has on average 13.51 trust groups.
As shown in Table~\ref{tab:trust-groups}, the mean size of these trust groups is 1.50.
We observe, that the size of trust groups differ between our conducted investigations.
This indicates that some trust relations identified via TLS certificates do not enable the resumption of TLS session across the corresponding hostnames.
Subsequently, we study these both sets of trust-relation by computing their union.
We notice that also a significant share of trust relations determined via successful TLS resumptions is uncovered by the trust relations identified via TLS certificates.
Considering the trust relations of both sets, we find a mean size of the trust groups per Alexa Top~1K Site of 2.42.

\begin{figure}[tbp]
\centering
\includegraphics[width=0.47 \textwidth]{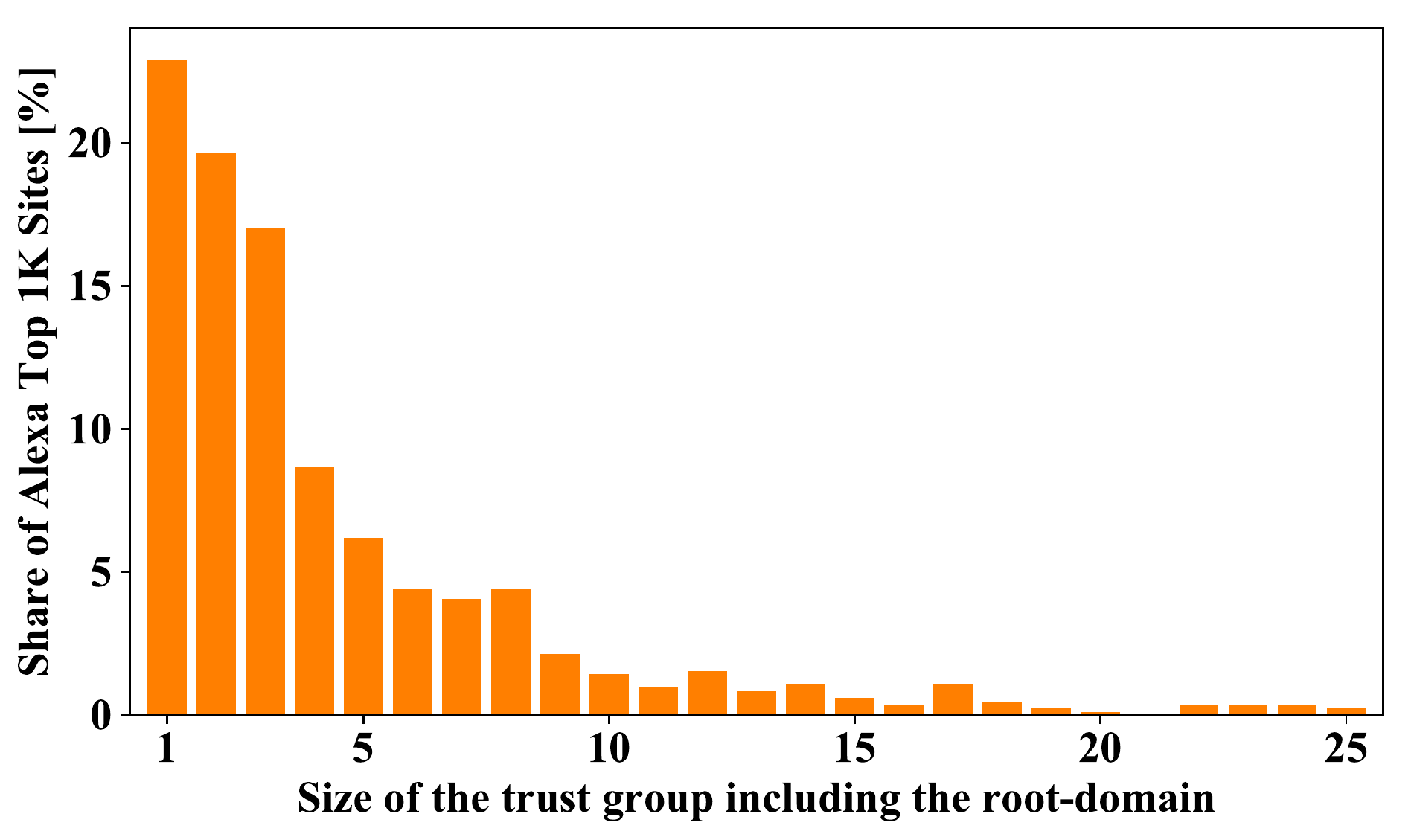}
\caption{Share of Alexa Top~1K Sites is plotted over the size of the trust group that includes the root-domain. Note, that this plot is cut off at a trust group size of 25.}
\label{fig:root_trustgroup_size}
\end{figure}

Figure~\ref{fig:root_trustgroup_size} provides additional insights into the distribution of the group size.
It plots the share of Alexa Top~1K Sites in dependence of the size of trust groups including the root-domain.
We find, that 77.12\% of the root-domains have a trust-relation to another hostname within their domain tree.
Furthermore, 90.94\% of the root-domains have a trust-relation to less than eleven hostnames within their domain tree.
Popular configurations have one or two trust relations for the root-domain, which represent 19.66\% and 17.04\% of the Alexa Top~1K Sites respectively.
Furthermore, we observe that 40 websites within the Alexa Top~1K Sites are served via their root-domain and do not require an additional TLS connection to a subdomain for their retrieval.

\subsection{Simulating TLS resumption across hostnames}

In this section, we study the real-world impact of TLS resumption across hostnames.
For this purpose, we conduct a simulation based on the loading behavior of the Alexa Top~1K Sites (see Section~\ref{sec:loading_behavior}) and the measured trust relations within their domain trees (see Section~\ref{sec:trust_relations}).
We start by investigating the performance impact of our proposal when loading a single website.
Subsequently, we extend this scenario by analyzing the visits to different websites. 

\subsubsection{Visiting a single website}\label{sec:single_website}

Our proposal is capable to significantly accelerate the first visit of an average website.
To substantiate this claim, we provide simulation results for the Alexa Top~1K Sites in this section.

Our simulation assumes the domain trees and trust groups as described in Section~\ref{sec:loading_behavior} and~\ref{sec:trust_relations}.
Furthermore, we assume that a full TLS handshake to any member of a trust group enables us to subsequently establish resumed connections to all other hostnames within the same trust group. In the real world that might not be always the case, as multiple TLS connections might be established in parallel, resulting in more full handshakes.

Using the proposed TLS extension, we can reduce the number of required full handshakes to download a website by 58.75\% on average.
Absolute numbers are provided in Table~\ref{tab:fullhandshake_cross_host} and indicate that our approach converts on average 11.89 full to resumed handshakes.
Furthermore, Table~\ref{tab:fullhandshake_cross_host} indicates that the traditional TLS resumption mechanism cannot reduce the number of full handshakes to different hostnames.
However, the traditional TLS resumption is still beneficial when multiple TLS connections to the same hostname are required for loading the website.

Assuming that each resumed TLS connection saves approximately 6~ms of CPU time of each peer (see Table~\ref{tab:CPU_time_comparison}), then our proposal reduces the required CPU time to load a website on average by 71.34~ms.
Assuming a full handshake to require 8~ms of CPU time per peer, then our proposal saves 44.06\% of the CPU time to load an average website initially.

 \begin{table}[htbp]
   \caption{Mean number of required full TLS handshakes to different hostnames to download a website of the Alexa Top~1K list for the first time. Note, that within this simulation trust relations between hostnames are assumed based on the union of presented TLS certificates and practical TLS resumptions.}\label{tab:fullhandshake_cross_host}
 \centering
 \begin{tabular}{ccc}
 \toprule
\makecell{Without TLS\\resumption} & \makecell{With TLS\\resumption}& \makecell{With TLS resumption\\across hostnames}\\
 \midrule
20.24 & 20.24 & 8.35 \\
 \bottomrule
 \end{tabular}
 \end{table}
 
\begin{figure}[tbp]
\centering
\includegraphics[width=0.47 \textwidth]{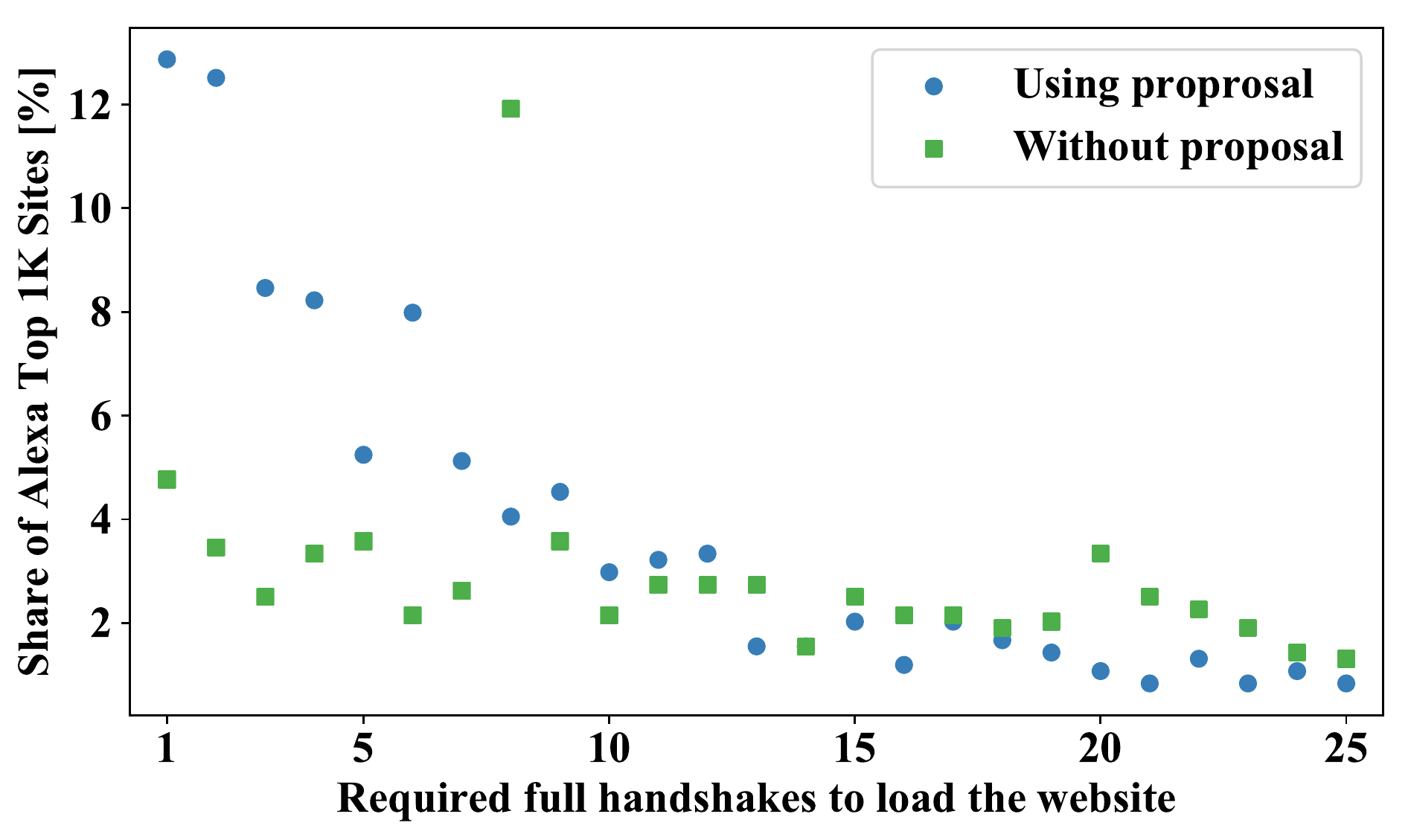}
\caption{This plot shows the share of Alexa Top~1K Sites over the number of required full TLS handshakes to retrieve the website. The blue circles mark the values considering TLS resumption across hostnames, while the green squares plot the current default (see Figure~\ref{fig:required_handshakes}). Note, that this plot is cut off at 25 full handshakes.}
\label{fig:trustgroup_number}
\end{figure}

 Figure~\ref{fig:trustgroup_number} shows the share of Alexa Top~1K Sites in dependence on the number of required full TLS handshakes to retrieve the website.
The results of our simulation are shown as blue circles and the default website loading behavior is marked with green squares (see Figure~\ref{fig:required_handshakes}).
We find that the share of websites that require less full handshakes significantly increased when TLS resumption across hostnames is used.
For example, the share of Alexa Top~1K Sites requiring only one full TLS handshake is 12.87\% when using TLS resumption across hostnames and 4.76\% without considering our proposal.
In total, the share of Alexa Top~1K Sites that require less than five full handshakes is now 42.07\% compared to 14.09\% without our TLS extension.
Furthermore, 95.95\% of the Alexa Top~1K Sites require less than 26 full TLS handshakes following our approach compared to 73.3\% of these websites without enabling TLS resumption across hostnames. 
 
In the following, we investigate the number of required sequential TLS full handshakes to retrieve an average Alexa Top~1K Site.
As shown in Table~\ref{tab:full-handshake-request-path}, the longest path of full TLS handshakes within the domain tree of an average Alexa Top~1K Site is 4.04, independently of the support for traditional TLS resumption.
Our proposal significantly reduced the number of TLS full handshakes required to retrieve an average website.
The longest path of full TLS handshakes decreased to 2.46. Thus, to establish all TLS connections required to retrieve a website on average requires only 2.46 full TLS handshakes and 1.58 resumed handshake instead of 4.04 full handshakes.
Equation~\ref{eq:delta_connect} allows the computation of $\Delta_{connect}$, which we define as the reduced delay until all TLS connections of an average website are established using TLS resumption across hostnames instead of the current default website loading.
\begin{equation}\label{eq:delta_connect}
\Delta_{connect} =
  \begin{cases}
    1.58 * \Delta_{1RTT}       & \quad \text{ for 1-RTT resumptions}\\
   1.58 * \Delta_{0RTT}  & \quad \text{ for 0-RTT resumptions}
  \end{cases}
\end{equation}

With respect to relative numbers, we achieve the most significant improvement for short network latencies.
To establish all connections for an average website via full TLS~1.3 handshakes requires $4.04*29.17~ms=117.85~ms$ for a network latency of approximately 0.3 ms.
TLS resumption across hostnames allows to save $1.58*22.83~ms=36.07~ms$ in this scenario, leading to a performance gain of 30.6\% until all connections are established.

 \begin{table}[htbp]
   \caption{Mean length of the longest path of required full TLS handshakes to different hostnames to retrieve an average website of the Alexa Top~1K list for the first time. }\label{tab:full-handshake-request-path}
 \centering
 \begin{tabular}{ccc}
 \toprule
\makecell{Without TLS\\resumption} & \makecell{With TLS\\resumption}& \makecell{With TLS resumption\\across hostnames}\\
 \midrule
4.04 & 4.04 & 2.46 \\
 \bottomrule
 \end{tabular}
 \end{table}

\begin{figure}[tbp]
\centering
\includegraphics[width=0.47 \textwidth]{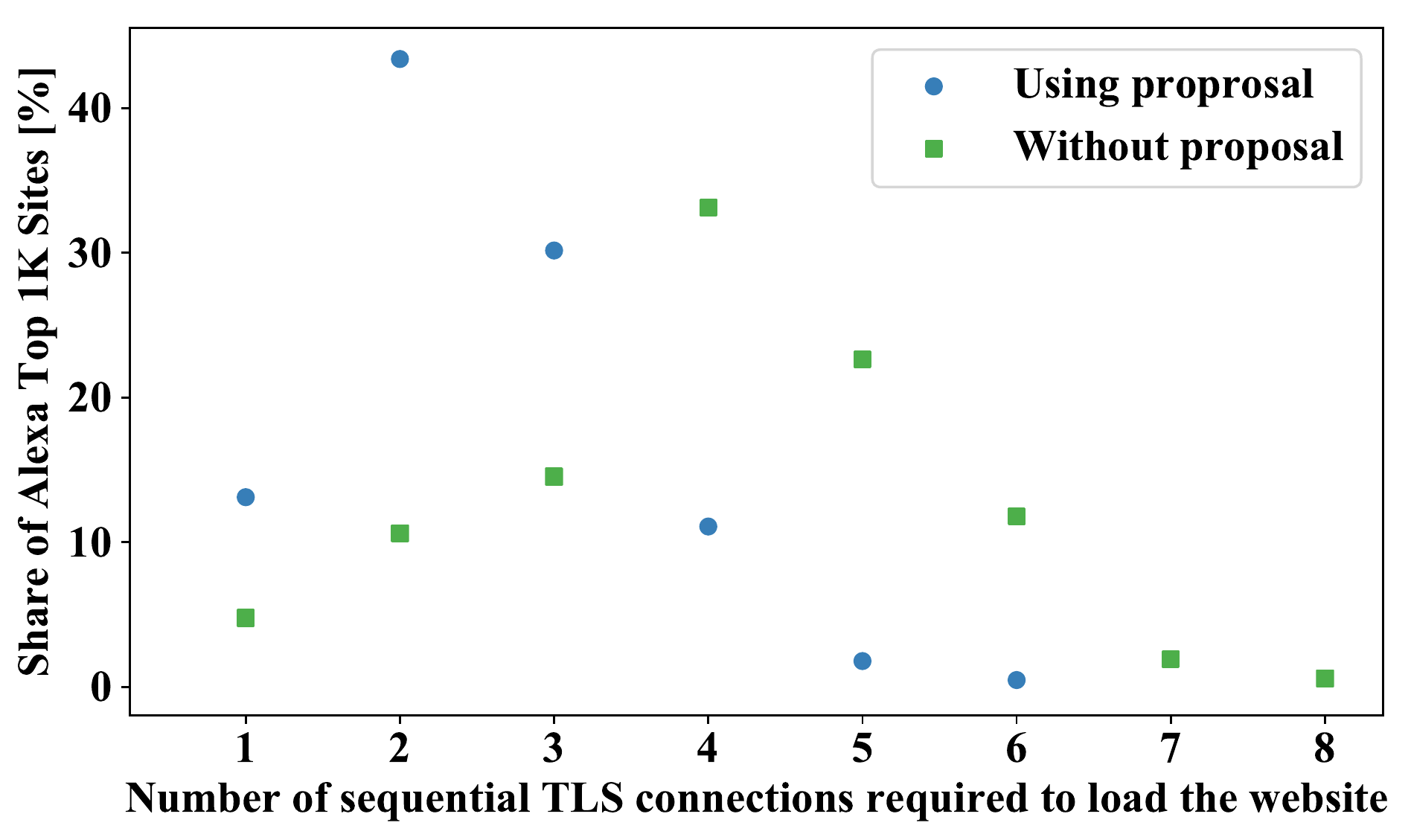}
\caption{This plot shows the share of Alexa Top~1K Sites over the number of required sequential TLS connections to load the respective website. The blue circles represent the values considering TLS resumption across hostnames, while the green squares plot the current default (see Figure~\ref{fig:longest_fullhandshake_path}).}
\label{fig:longest_fullhandshake_cross_host_path}
\end{figure}

Figure~\ref{fig:longest_fullhandshake_cross_host_path} provides additional insights on the number of required sequential TLS full handshakes to retrieve an average Alexa Top~1K Site.
Comparing our simulation results plotted as blue circles to the default website loading behavior represented as green squares (see Figure~\ref{fig:longest_fullhandshake_path}), we find that TLS resumption across hostnames significantly reduces the number of sequential TLS full handshakes.
For example, the number of Alexa Top~1K Sites require less than three sequential TLS full handshakes for their retrieval has approximately quadrupled by using TLS resumption across hostnames. 
Furthermore, the share of Alexa Top~1K Sites that require more than five sequential full TLS handshakes decreased from 36.95\% to 2.26\%.

\subsubsection{Visiting different websites}

TLS resumptions are not just feasible within the context of a single website.
For example, if different websites require connections to the same hostname, then a sequential visit to these websites makes TLS resumptions feasible.
In this section, we deploy the same assumptions as in Section~\ref{sec:single_website} to study visits to different websites.
However, we consider only trust relations from the TLS certificates for the simulation as an experimental resumption test between all 16\,984 hostnames present in the Alexa Top~1K sites would result in almost 290~million connection attempts to a rather small group of servers.

In our test scenario, we measure the average of the required number of full handshakes to different hostnames, when visiting different websites within the Alexa Top list.
Table~\ref{tab:fullhandshake_cross_host_two_sites} summarizes these results for the Alexa Top~100 and the Alexa Top~1K Sites.
We find, that the traditional TLS resumption converts about 2.5\% of the full handshakes to resumed connection establishments in the Alexa Top~100 test scenario.
In the same scenario, our proposed approach converts 64.2\% of the full handshakes to resumed connection.
In the larger Alexa Top~1K simulation, we find that traditional TLS resumption allows to reduce the number of full TLS handshakes by 4.1\%, while our approach reduces this number by 55.5\% less full TLS handshakes at the same time.
These results substantiate, that TLS resumption across hostnames significantly improves performance for the encrypted web.

 \begin{table*}[htbp]
   \caption{Mean number of required full TLS handshakes to different hostnames to download two websites of a given Alexa Top list. Note, that within this simulation trust relations between hostnames are assumed based on the presented TLS certificates.}\label{tab:fullhandshake_cross_host_two_sites}
 \centering
 \begin{tabular}{lccc}
 \toprule
\multicolumn{1}{c}{Alexa Top list} & \makecell{Without TLS resumption} & \makecell{With TLS resumption}& \makecell{With TLS resumption\\across hostnames}\\
 \midrule
\makecell{Alexa Top 100}& 25.98 & 25.34 & 9.30 \\
\makecell{Alexa Top~1K}& 41.78 & 40.04 & 18.58 \\
 \bottomrule
 \end{tabular}
 \end{table*}

\paragraph{Summary}
In summary, our results indicate that resumed handshakes have a significantly reduced delay and computational overhead compared to full TLS handshakes.
Moreover, we find that the loading of an average website requires more than 20 TLS connections to separate hostnames, which tend to have a trust relation with each other.
To reduce the overhead caused by the TLS connection establishment, TLS resumption across hostnames provides an efficient mechanisms.
Thus, the average website can be retrieved with 44.06\% less CPU time consumed.
Furthermore, the establishment of the required connections to retrieve the website can be accelerate with up to 30.6\% based on our TLS~1.3 test measurements.

\section{Discussion}\label{sec:Discussion}

In this section, we discuss the security and privacy impact of the proposed TLS extension.
Subsequently, we review developments on the Internet that may affect the impact and benefit of our proposal. 

\subsection{Security considerations}

The authentication of the server's identity is a critical part of the TLS handshake.
In the full TLS handshake, the client validates the server's identity based on public-key cryptography.
Thus, the server is required to present a valid certificate containing a public key that confirms the claimed identity.
Furthermore, the client validates that the server can generate a fresh signature with the private key corresponding to the presented certificate/public key.

Within a resumed TLS handshake, these computationally expensive public-key operations are omitted.
The server is authenticated by its knowledge of a cryptographic secret related to the original TLS session, which allows the server to decrypt parts of the resumption handshake.
This practice presents an indirect authentication of the server's identity because the client does not validate the server's certificate and the server's possession of the corresponding private key within the resumed TLS session.
Thus, resumption handshakes require the client to trust the correctness of the server authentication during the original session.

In the scenario of TLS resumption across hostnames, the risk arises that an attacker exploits this weaker validation of the server's identity during the resumption handshake for impersonification attacks.
For example, the client connects with a full TLS handshake to a legitimate but malicious SNI.
The corresponding malicious server then advertises session resumption to an illegitimate SNI that would not withstand a validation of its identity by the client during a full TLS handshake. By using a resumption handshake to connect to this illegitimate SNI, the client may successfully establish a TLS connection without noticing the insufficient authentication of the server's identity.
 
To avoid such attacks, resumption handshakes need to be restricted to SNI values that are valid with respect to the server certificate presented in the original session.
Note, that this restriction is already a requirement in the specification of TLS~1.3~\cite{rfc8446}.

\subsection{Privacy considerations}

Session resumption mechanisms in TLS~1.3 allow tracking a user across several visits to the same hostname~\cite{sy2018tracking}.
TLS~1.3 uses unique pre-shared key (PSK) identities for session resumption that can be linked to a specific user.
Thus, every time a user presents a cached PSK identity during a connection attempt, this new connection can be linked to the original connection where that unique PSK was issued by the server to the client.
By enabling session resumption across several hostnames, the operators of these hostnames can link user visits to any of these hostnames.
For that, the operators of the contacted hostnames need to share logs with each other.
These logs contain an entry on the freshly issued PSKs to a user, details on the user activities, and if applicable information on the presented PSK by the client during the resumption handshake.
Thus, by using the currently available session resumption mechanisms of TLS and when extending this across hostnames the problem of illegitimate user tracking across hostnames is facilitated.

User tracking across hostnames is a long established privacy problem on the Internet~\cite{bujlow2017survey}.
Assuming that two different websites want to track a specific user, e.g., via HTTP cookies or via TLS session resumption mechanisms, they can use a simple web link with a unique URL. If the user follows that link, then the operators of the hostnames associate the user in their logs with the observed web link.
By sharing their logs and matching the included web links the operators can extract the user behavior across the respective hostnames.
This example indicates that simple web links are sufficient to enable user tracking across hostnames and our proposed approach does not introduce completely new privacy problems for the user.

The average performance of TLS connection establishments depends on the ratio of resumed handshakes per full handshakes.
The higher this ratio is, the lower is the overhead of the TLS connection establishment.
The lifetime of the TLS session resumption mechanisms impacts this performance for a given browsing session, because a shorter lifetime leads to less resumed handshakes per full handshakes.
Furthermore, this lifetime presents an upper bound for the feasible tracking period by the session resumption mechanism~\cite{sy2018tracking}.
By using resumption handshakes across hostnames, we yield the same ratio of resumed handshakes per full handshakes within a shorter lifetime of the session resumption mechanism.
For example, we consider a website that requires TLS connections to two hostnames that support session resumption between each other. By using our approach, we can directly retrieve the website with a full and a resumed TLS connection leading to a ratio of one.
Without the usage of resumption handshakes across hostnames, it requires two retrievals of the same website to yield the same ratio of a resumed handshake per full handshake.
Thus, our approach allows setting the lifetime of the session resumption mechanism to a single website visit to achieve the same performance as the two website retrievals following the current practice of session resumption.
A short lifetime of the resumption mechanism restricts the feasible tracking periods and therefore our approach can lead to improvements for the user's privacy without impacting the performance of the TLS connection establishment.

\subsection{Chances and limitations}

In this section, we review the growing TLS adoption on the web as this development positively contributes to the real-world impact of our proposed TLS extension.
Subsequently, we investigate the adoption of TLS~1.3 0-RTT resumption handshake by popular websites.
This handshake mode does not inherently protect against replay attacks, which restricts its use cases compared to 1-RTT resumption handshake.
If the 0-RTT handshake mode exhibits a lower adoption compared to the 1-RTT resumption mode, then this practice limits the feasible performance optimizations of TLS~1.3 resumption handshakes across hostnames.

\subsubsection{Growing TLS adoption on the web}

\begin{figure}[t]
\centering
\includegraphics[width=0.47 \textwidth]{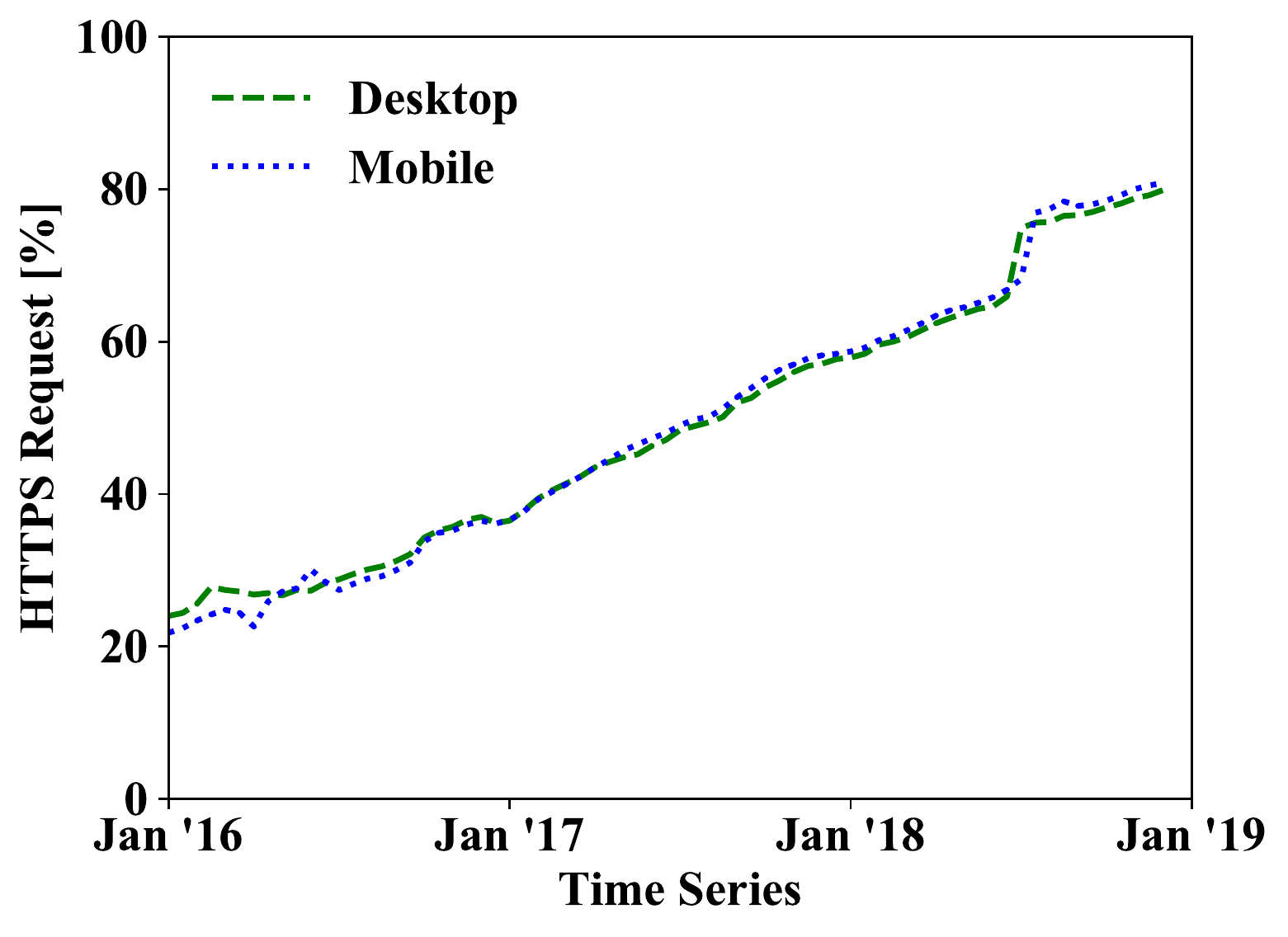}
\caption{Percent of all requests indicating TLS support is plotted over a time series from January 2016 to December 2018.}
\label{fig:growing_adoption}
\end{figure}

The adoption of TLS has significantly increased during the last years~\cite{felt2017measuring}.
To substantiate this statement, Figure~\ref{fig:growing_adoption} shows of the share of encrypted HTTPS requests over a three years' time series.
 This data is collected by regularly retrieving websites using desktop (dashed line) and mobile (dotted line) browsing environments.
 For further details on the methodology of these web scans, we refer the reader to~\cite{HTTP_Archive}.
 
As shown in Figure~\ref{fig:growing_adoption}, the share of HTTPS request on the web increased by more than 50\% during the plotted period.
Research work~\cite{felt2017measuring} expects even further growth of the TLS deployment on the web.
Prior research\cite{sy2018tracking} on TLS session resumption indicates that about 96\% of the websites supporting TLS do also support a session resumption mechanism.
Thus, we expect that the increased deployment of TLS on the web will raise the absolute number of hostnames that support session resumption.

\subsubsection{Measuring the adoption of TLS~1.3 handshake modes}

Besides the full handshake mode, TLS~1.3 provides the option to support 1-RTT and 0-RTT resumption handshakes.
We collected data on the Alexa Top Million Sites~\cite{Alexa} to investigate the adopation of the different TLS~1.3 handshake modes.
We scanned these websites on January 25, 2019 using the library boringssl~\cite{BoringSSL}.
For that, we first established a TLS~1.3 connection via a full TLS handshake. If this connection establishment succeeded, then we directly tried to establish connections via 1-RTT and 0-RTT resumption handshakes.

Table~\ref{tab:alexa_lists} summarizes our findings. 80\,799 websites within the Alexa Top Million list support the TLS~1.3 full handshake. 87.6\% of these websites also support session resumption via 1-RTT handshakes. However, within the Alexa Top 100K this share reaches 96.9\%. Thus, our results indicate widespread support for session resumption by websites supporting TLS~1.3.
Furthermore, our results indicate that 491 websites among the Alexa Top Million list support the 0-RTT handshake mode.
This represents a share of 0.6\% of the respective websites within the Alexa Top Million list that support TLS~1.3. 
However, within the Alexa Top~1K this share raises to 7.1\% with nine websites out of 127 websites supporting TLS~1.3.
Overall, the support of 0-RTT handshakes is within our sample lists, at least one magnitude lower compared to the support of 1-RTT resumption handshakes.

 \begin{table}[t]
   \caption{Websites with TLS~1.3-support in Alexa Top lists. Additionally the support of the 1-RTT and 0-RTT session resumption (SR) handshake is indicated.}
  \label{tab:alexa_lists}
 \centering
 \begin{tabular}{lrrr}
 \toprule
\multicolumn{1}{c}{Alexa Top lists} & \multicolumn{3}{c}{\makecell{Share of websites supporting TLS~1.3}} \\
 \cmidrule(lr){2-4}
& \multicolumn{1}{c}{Full mode} & \multicolumn{1}{c}{1-RTT SR}& \multicolumn{1}{c}{0-RTT SR}\\
 \midrule
Alexa Top 10&     10.0\%&10.0\%&0.00\%\\
 Alexa Top 100&  9.0\%&8.0\%&0.00\%\\
 Alexa Top~1K&    12.7\%&11.6\%&0.90\%\\
 Alexa Top 10K&  13.1\%&12.6\%&0.44\%\\
 Alexa Top 100K& 9.2\%&8.9\%&0.19\%\\
 Alexa Top 1M &    8.1\%&7.1\%&0.05\%\\
 \bottomrule
 \end{tabular}
 \end{table}

In total, we find our measurement investigates an early stage of the deployment of TLS~1.3 on the web.
At the time of our scan, still many server vendors and CDNs did not support TLS~1.3~\cite{istlsfastyet}.
Furthermore, some server vendors and CDNs do support TLS~1.3 except for TLS~1.3 0-RTT resumption handshakes~\cite{istlsfastyet}.
We expect that the number of websites supporting TLS~1.3 and their resumption handshakes will raise during the next years as more server vendors and CDNs will support this new TLS version by default.
We assume that the lower deployment of TLS~1.3 0-RTT resumption handshakes is also caused by the lower security guarantees compared to the 1-RTT resumption case.
Thus, we expect this handshake mode will remain less widely adopted compared to the 1-RTT handshake mode. 


\section{Related Work}\label{sec:Related}

To the best of our knowledge, we are the first to investigate the performance benefit of TLS resumption across hostnames.
However, prior work~\cite{sy2018tracking, Springall:2016:MSH:2987443.2987480} did already report the sharing of TLS state across hostnames, which is a pre-requisite for resuming a TLS session with another hostname.

Furthermore, also RFC~8446~\cite{rfc8446} briefly discusses the opportunity of performance enhancements based on TLS resumption across hostnames.
This RFC concludes that TLS resumption across hostnames is possible within certain security limitations that lead to a high failure rate for resumption attempts.
However, this RFC did not consider the option of a TLS extension to inform the client about other hostnames that are capable to resume the respective session.
We argue that our proposed TLS extension minimizes this failure rate, as the client will only attempt a resumption handshake with hostnames directly recommended by the server during the original TLS session.

\section{Conclusions}\label{sec:Conclusion}

A TLS extension supporting clients to conduct TLS resumption across hostnames is overdue.
Our evaluations indicate, that the proposed TLS extension yields great performance optimizations for clients and servers on the existing web without affecting the user's privacy and communication security.

\bibliographystyle{ACM-Reference-Format}
\bibliography{sample-bibliography}

\end{document}